\documentclass[twoside]{article}
\usepackage{fleqn,espcrc2,epsfig}

\def\Journal#1#2#3#4{{#1} {\bf #2}, #3 (#4)}
\def\NJP{\em New J. Phys.}
\def\NIMA{{\em Nucl. Instrum. Methods} A}
\def\NPB{{\em Nucl. Phys.} B}
\def\PLB{{\em Phys. Lett.}  B}
\def\PRL{\em Phys. Rev. Lett.}
\def\PRC{{\em Phys. Rev.} C}
\def\PRD{{\em Phys. Rev.} D}
\def\PPNP{\em Prog. Part. Nucl. Physics}
\newcommand{\mus}{\mbox{$\mu$s}}
\newcommand{\us}{\mbox{$\mu$s}}
\newcommand{\C}{\mbox{$^{12}$C}}

\newcommand{\Cd}{\mbox{$^{13}$C}}

\newcommand{\Fe}{\mbox{$^{56}$Fe}}
\newcommand{\N}{\mbox{$^{12}$N}}
\newcommand{\ep}{\mbox{e$^{+}$}}

\newcommand{\el}{\mbox{e$^{-}$}}
\newcommand{\pos}{\mbox{e$^{+}$}}
\newcommand{\mum}{\mbox{$\mu^{-}$}}
\newcommand{\mup}{\mbox{$\mu^{+}$}}
\newcommand{\pim}{\mbox{$\pi^{-}$}}
\newcommand{\pip}{\mbox{$\pi^{+}$}}
\newcommand{\numu}{\mbox{$\nu_{\mu}$}}
\newcommand{\numub}{\mbox{$\bar{\nu}_{\mu}$}}
\newcommand{\nue}{\mbox{$\nu_{e}$}}
\newcommand{\nueb}{\mbox{$\bar{\nu}_{e}$}}
\newcommand{\mupdecay}{\mbox{\mup\ $\rightarrow\:$ \pos $\!$ + \nue\ + \numub}}

\newcommand{\pipmup}{\mbox{\pip $\rightarrow\:$ \mup + \numu}}
\newcommand{\pipmupx}{\mbox{\pip $\rightarrow\:$ \mup + $X$}}

\newcommand{\CCprot}{\mbox{p\,(\,\nueb\,,\,\ep\,)\,n }}
\newcommand{\excl}{\mbox{\C\,(\,\nue\,,\,\el\,)\,\N$_{\rm g.s.}$}}
\newcommand{\numunue}{\mbox{\numu $\rightarrow\,$\nue }}
\newcommand{\numubnueb}{\mbox{\numub $\rightarrow\,$\nueb }}

\newcommand{\NCL}{\mbox{90\%\,CL}}
\newcommand{\NnCL}{\mbox{95\%\,CL}}
\newcommand{\Dm}{\mbox{$\Delta m^2$}}
\newcommand{\sit}{\mbox{$\sin ^2(2\Theta )$}}
\newcommand{\eVc}{\mbox{eV$^2$/c$^4$}}
\newcommand{\Gdng}{\mbox{Gd\,(\,n,$\gamma$\,)}}
\newcommand{\pnd}{\mbox{p\,(\,n,$\gamma$\,)\,d}}

\title{Latest Results of the KARMEN2 Experiment}

\author{K. Eitel\address{Institut f\"ur Kernphysik, 
        Forschungszentrum Karlsruhe,
        P.O. Box 3640, D-76021 Karlsruhe, Germany\\
	e-mail: klaus@ik1.fzk.de}
	representing the KARMEN collaboration \cite{karmen} }
       
\begin{document}

\begin{abstract}
The neutrino experiment KARMEN at the beam stop neutrino source
ISIS investigates the oscillation channel \numubnueb\ in the appearance 
mode by looking for \CCprot\ reactions. An analysis of data
collected from February 1997 through March 2000 with the KARMEN2 
experimental setup reveals 11 candidate events in good agreement with 
the background expectation of $12.3\pm 0.6$ events. Hence, there is no 
indication of an oscillation signal. A maximum likelihood analysis 
of the data leads to an upper limit (at 90\% confidence level)
 for the mixing angle of 
 $\sit < 1.3\cdot 10^{-3}$ at large \Dm\ and $\Dm < 0.049$\,\eVc\ for
 $\sit=1$.

The anomaly in the time spectrum of events induced by \nue\ and \numub\
seen in the KARMEN1 data could not be confirmed with the KARMEN2 data.
\vspace{1pc}
\end{abstract}

\maketitle

\section{THE EXPERIMENT}

The KARMEN experiment
is performed at the neutron spallation facility ISIS of the Rutherford 
Appleton Laboratory. Neutrinos are produced by stopping 800\,MeV protons 
in a massive beam stop target, thereby producing pions. 
The \pim\ are absorbed by the target 
nuclei whereas the \pip\ decay at rest (DAR). Muon neutrinos \numu\ therefore 
emerge from the decay \pipmup . The low momentum \mup\ are also stopped 
within the massive target and decay via \mupdecay .
Because of this \pip -\mup -decay chain at rest ISIS represents a $\nu$-source
with identical intensities for \numu , \nue\ and \numub\ emitted 
isotropically ($\Phi_{\nu}=6.37\cdot 10^{13}$\,$\nu$/s per flavor for 
a proton beam current $I_p=200$\,$\mu$A). 
  \begin{figure}[hbt]
  \centerline{\epsfig{figure=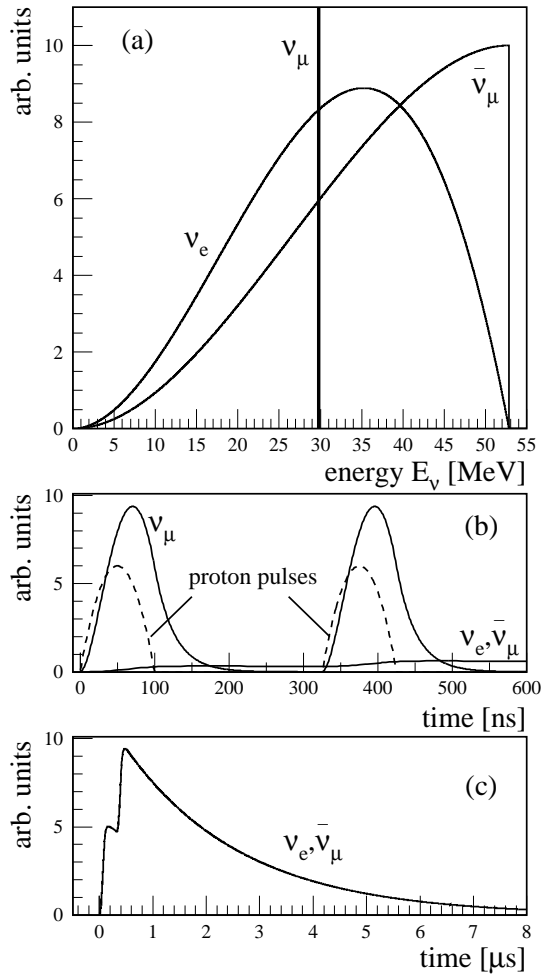,width=7cm}}
  \caption{Neutrino energy spectra (a) and production times of \numu\ (b)
	   and \nue ,\numub\ (c) at ISIS.}
  \label{isis_nu}
  \end{figure}
There is a minor fraction of \pim\ decaying in flight (DIF) with the following 
\mum\ DAR in the target station which again is suppressed by muon capture of 
the high $Z$ material of the spallation target. This decay chain leads to 
a very small contamination of $\nueb/\nue < 6.2\cdot 10^{-4}$ \cite{bob}, which
is further reduced by software cuts.

The energy spectra of the $\nu$'s are well defined due to the DAR
of both the \pip\ and \mup\ (Figure~\ref{isis_nu}a). The \numu 's from 
\pip --decay are monoenergetic with E(\numu)=29.8\,MeV, the continuous 
energy distributions of \nue\ and \numub\ up to $52.8$~MeV can be calculated 
using the V--A theory.
Two parabolic proton pulses of 100\,ns base width and a gap of 225\,ns are 
produced with a repetition frequency of 50\,Hz. The different lifetimes of 
pions ($\tau$\,=\,26\,ns) and muons ($\tau$\,=\,2.2\,$\mu$s) allow a clear 
separation in time of the \numu -burst (Figure~\ref{isis_nu}b) from the 
following \nue 's and \numub 's (Figure~\ref{isis_nu}c). The accelerator's 
duty cycle of $10^{-5}$ allows effective suppression of cosmic induced
background.

The neutrinos are detected in a rectangular tank filled with 56\,t of a liquid 
scintillator \cite{detector}. The central scintillation calorimeter is
segmented into 512 optically individual modules.
The event position is determined by the hit module and 
the time difference of the PM signals at each end of this module. Gd$_2$O$_3$ 
coated paper within the module walls provides efficient detection of thermal 
neutrons due to the very high capture cross section of the \Gdng\ reaction
($\sigma \approx 49000$\,barn) in addition to the \pnd\ capture. 
The KARMEN electronics is synchronized to the 
ISIS proton pulses to an accuracy of better than $\pm 2$\,ns, so that the time 
structure of the neutrinos can be exploited in full detail.

A massive blockhouse of 7000\,t of steel in combination with a system 
of two layers of active veto counters provides shielding 
against beam correlated spallation neutron background, suppression
of the hadronic component of cosmic radiation as well as reduction of the flux
of cosmic muons. In 1996 an additional third 
veto counter system with a total area of 300\,m$^2$ was installed within 
the 3\,m thick roof and the 2--3\,m thick walls of the iron shielding 
\cite{drexlin} (KARMEN2 experimental configuration). By detecting muons which 
pass through the steel at a distance of less than a meter from the main 
detector and therefore vetoing the successive energetic neutrons from muon
deep inelastic scattering, the main background for the \numubnueb\ oscillation
search could be reduced by a factor of 40 compared to the KARMEN1 setup.

\section{THE \numubnueb\ OSCILLATION SEARCH}

The signature for the detection of \nueb\ is a spatially correlated 
delayed coincidence of positrons from \CCprot\ with energies up to 
$E_{e^+}=E_{\nueb}-Q=52.8-1.8=51.0$\,MeV and 
$\gamma$ emission of either of the two neutron capture processes \pnd\ 
with one $\gamma$ of $E(\gamma)=2.2$\,MeV or \Gdng\ with 3 $\gamma$--quanta 
on average and a sum energy of $\sum E(\gamma)=8$\,MeV.
The positrons are expected in a time window of several \mus\ after
beam--on--target with a 2.2\,\us\ exponential 
decrease due to the \mup\ decay. The time difference between the \pos\
and the capture $\gamma$ is given by the thermalization, diffusion and capture
of neutrons, $\tau_n\approx 110$\,\us .

The raw data investigated for this oscillation search were recorded in the
measuring period  of February 1997 through March 2000 which corresponds to
7160\,C protons on target. A positron candidate is accepted only if there 
is no previous activity in the central detector nor in the two innermost
veto counters up to 24\,\us . The required cuts in energy and time 
for the prompt ($p$) event are:
$0.6 \le t_p \le 10.6$\,\us , $16.0\le E_p \le 50.0$\,MeV. The cuts on the 
delayed expected neutron event are as 
  \begin{figure}[hbt]
  \centerline{\epsfig{figure=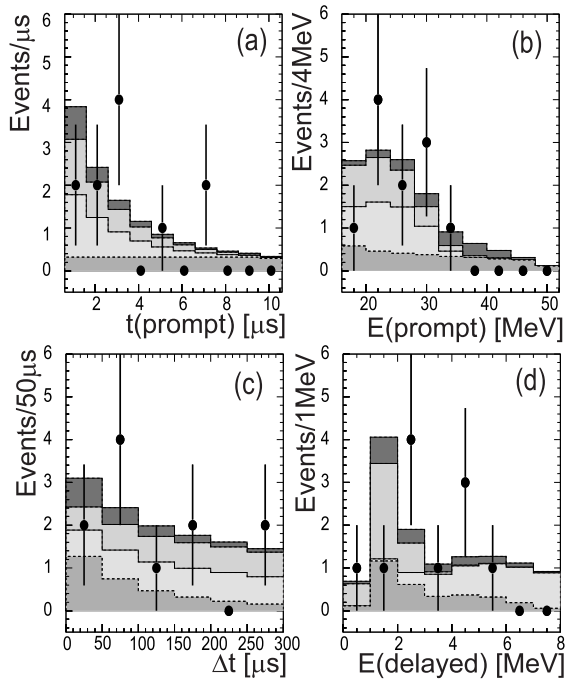,width=7.5cm}}
  \caption{Spectra of the 11 candidate sequences after applying all cuts.
	Also shown are the background contributions (from bottom to top):
	cosmic background, (\el ,\ep) from \excl , $\nu$ induced random 
	coincidences and ISIS \nueb\ contamination.}
  \label{4erplot}
  \end{figure}
follows: $5.0\le t_d - t_p \le 300$\,\us , $E_d \le 8$\,MeV and a
volume of 1.3\,m$^3$ for the spatial coincidence. 
Applying all cuts, the total background expectation amounts to $12.3\pm0.6$
sequences where the individual background sources are events induced by 
cosmic muons, \excl\ sequences, $\nu$-induced accidental coincidences and 
(\pos ,n) sequences from the intrinsic ISIS \nueb\ contamination. 

In Table~\ref{backgrd} the individual contributions of the above described
background sources are summarized. 
  \begin{table}[hbt]
  \caption{Expected sequences from different background components within
           the evaluation cuts specified in the text. Last row: oscillation
	   expectation for maximal mixing.}
  \begin{tabular}{lr} \hline
        background contribution                 & events  \\ \hline
        (\el ,\ep) from \excl\ 			&  3.9$\pm$0.5   \\
        $\nu$ induced random coincidences       &  3.5$\pm$0.3   \\ 
        ISIS \nueb\ contamination               &  1.7$\pm$0.2   \\ 
        cosmic induced sequences                &  3.2$\pm$0.2   \\ \hline
        total background                        & 12.3$\pm$0.6   \\ \hline
        \nueb\ signal for $\sit=1$              & 2442$\pm$269   \\ \hline
  \end{tabular}
  \label{backgrd}
  \end{table}
The last row shows the expectation of (\pos ,n) sequences from oscillations 
assuming maximal mixing and $\Dm \ge 100$\,\eVc .
The background components are determined precisely during the normal
measurements: The $\nu$ induced backgrounds are measured with KARMEN in 
different energy and delayed time windows, the cosmic background is measured
with high statistics in the long pre-beam time window. Only the small \nueb\
contribution of the intrinsic source contamination has to be simulated.
In addition, the capability of the KARMEN experiment to identify neutrino
induced events is constantly monitored by measuring neutrino--nucleus 
interactions on \C\ via neutral and charged current reactions.

Analysing the data results in 11 sequential events which satisfy all 
conditions (see Figure \ref{4erplot}). This number is in good agreement with
the total background expectation.
Applying a Bayesian renormalisation of the physically 
allowed region to the experimental result near a boundary ($N(osc)\ge 0$, 
11 measured events with $12.3\pm 0.6$ background events expected),
an upper limit of $N(osc)<6.3$ at \NCL\ can be extracted. 
An identical procedure had been applied to the initial KARMEN2 data, when
the evaluation of a potential oscillation signal was based on a pure counting 
experiment due to the very small statistics of the background \cite{nu98}.

However, with more data and spectral information,
a much better evaluation method is a maximum likelihood analysis.
Such a likelihood analysis to extract a possible \numubnueb\ signal from
these 11 sequences makes use of the precise spectral knowledge of all 
background sources and a detailed MC description of the oscillation signature 
in the detector. The likelihood function $L$ defined as
$$ L = \prod_{i=1}^N f(\vec{x}_i,\Dm ,\sit ) $$ is optimised with respect to
the free parameters \Dm\ and \sit . The probability density function $f$ is
calculated for each of the $N$ event sequences from the parameters
 $\vec{x}=(E_{prompt},E_{delayed},t_{prompt},\Delta t,\Delta \vec{r})$ where
 $\Delta t$ and $\Delta \vec{r}$ denote the delayed spatial coincidence.
The analysis results in a best fit value of oscillation
events $N(osc)=0$ within the physically allowed range of parameters.
  \begin{figure}[hbt]
  \centerline{\epsfig{figure=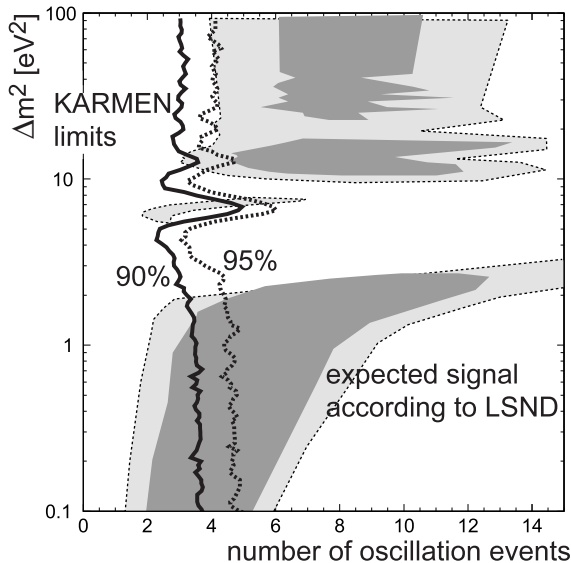,width=7.5cm}}
  \caption{KARMEN exclusion limits at \NCL\ and \NnCL\ in comparison with
	the range of expected oscillation events deduced from the LSND 
	evidence \protect\cite{mills}. The darker (lighter) region corresponds 
	to logarithmic likelihood values less than 2.3 (4.6) units below the 
	global maximum.}
  \label{os1}
  \end{figure}
As shown in Figure~\ref{os1}, an upper limit of 3.8 and 3.1 oscillation 
events for $\Dm<1$\,\eVc\ and $\Dm>20$\,\eVc , respectively, can be extracted 
at \NCL\ based on a complete
frequentist approach as suggested by \cite{cous}.

The limits at \NCL\ and \NnCL\ can be compared with the number of oscillation
events expected from the latest results of the LSND experiment \cite{mills}. 
In Figure~\ref{os1}, the calculated LSND signal strength within the 
KARMEN detector is shown. 
For large \Dm , even at \NnCL , all of the favored LSND signal range can be 
excluded by KARMEN2. At lower values of \Dm , a small fraction of the 
signal strength of LSND cannot be ruled out by KARMEN2.
Note that at high \Dm\ the band of expected events and the exclusion curve are 
almost parallel. With decreasing values of \Dm\ the expected signal strength
becomes smaller. This may be attributed, in part, to the larger source distance
for the LSND experiment ($\langle d_{LSND} \rangle \approx 30$\,m compared
to $\langle d_{KARMEN} \rangle \approx 17.7$\,m) but may also
reflect a smaller signal strength: If the LSND data evaluation yields an
oscillation probability $P(\numubnueb)$ which decreases with smaller values 
of \Dm\ --instead of remaining constant-- this would have the same effect.

Assuming maximal mixing ($\sit=1$), $2442\pm 269$ ($e^+$,n) sequences from 
oscillations with large \Dm\ were expected.
  \begin{figure*}[hbt]
  \centerline{\epsfig{figure=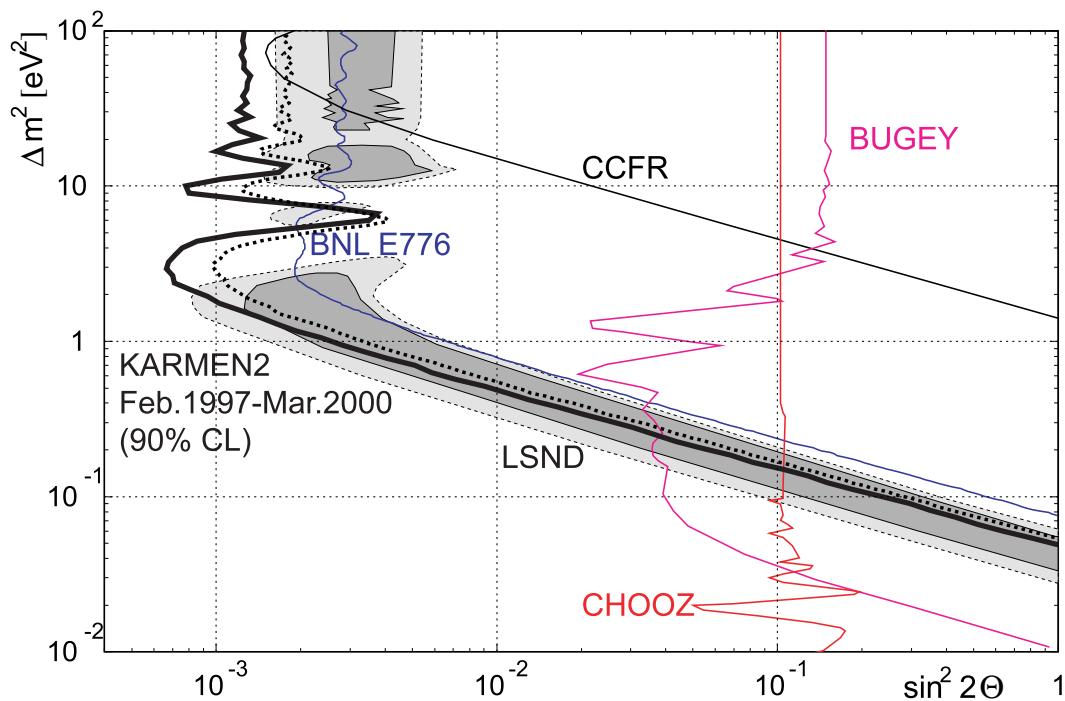,width=14cm}}
  \caption{KARMEN2 exclusion limit and sensitivity (dashed line) at \NCL\
	compared to other experiments: BNL E776 \protect\cite{bnl},
	CCFR \protect\cite{ccfr}, 
	BUGEY \protect\cite{bugey}, CHOOZ \protect\cite{chooz}  
	and the evidence for \numubnueb\
        oscillations reported by LSND \protect\cite{mills}. }
  \label{os2}
  \end{figure*}
Together with the limit of 3.1 events deduced in the unified frequentist
approach, this leads to an upper limit on the mixing amplitude of 
 $$\sit<1.3\cdot 10^{-3} \qquad (\NCL)$$ for $\Dm \ge 100$\,\eVc . 
The corrsponding exclusion curve
in (\Dm,\sit) is given in Fig.~\ref{os2}. Also shown are limits from
other experiments~\cite{bnl,ccfr,bugey,chooz} as well as the favored
regions from LSND based on a complete re-analysis of the entire 
1993-98 data set~\cite{mills}.
Again, at high \Dm , KARMEN excludes the region favored by LSND.
At low \Dm , KARMEN leaves some statistical space, but the reactor experiments
at Bugey and Chooz add stringent limits from the \nueb\ disappearance search.
Any exclusion curve or favored region is a compactification of a complex
statistical information, it reflects only a contour of the
twodimensional likelihood functions which have nontrivial properties like
multiple side maxima and non-parabolic shapes. Therefore, one needs
a quantitative statistical analysis of both LSND and KARMEN based on 
the detailed event-by-event information as demonstrated in \cite{NJP} for
preliminary data sets to deduce correct statements 
of compatibility or disagreement in terms of frequentist confidence regions.

Taking the LSND region published first \cite{atha} which corresponds to a 
stronger \numubnueb\ signal, there is a much stronger disagreement in the 
experiments' outcome: The actual \NCL\ KARMEN2 limit excludes the complete
parameter range favored by the 1993-95 LSND data.

One way of estimating the sensitivity of an experiment is to determine the
average limit on the oscillation parameter simulating a large number of 
experimental outcomes with the actual level of background events, but no 
oscillation signal. These samples are subsequently analysed with the same
maximum likelihood analysis used for the real data set.
The actual KARMEN limit is slightly better than its sensitivity of
$\sit=1.8\cdot 10^{-3}$ for large \Dm\ and almost identical at lower \Dm\
(see dashed line in Fig.~\ref{os2}).
Compared to the earlier results of KARMEN \cite{nu98}, this corresponds 
to an improvement of the sensitivity by a factor of $\approx 2.5$. 

KARMEN will continue to take data
for the \numubnueb\ search until the end of April 2001 with an anticipated 
sensitivity of $\sit=1.3\cdot 10^{-3}$ for large \Dm . Meanwhile,
the upcoming BooNE experiment at Fermilab is under construction.
Its sensitivity is expected to improve the final KARMEN sensitivity by
another factor of 2 \cite{bazarko} in the appearance mode \numunue\ with 
different systematics than KARMEN and LSND.

\section{THE TIME ANOMALY}

In the time window of 0.6-10.6\,\us\ after beam-on-target, 
reactions on \C , \Cd\ and \Fe\ induced by \nue\ and \numub\ are expected to
reflect the 2.2\,\us\ life time of the muons decaying in the target 
superimposed on a flat, cosmic induced background. However, in the KARMEN1
data there was a bump--like distortion of the spectrum 
at about 3.1--4.1\,\us\ after beam-on-target first reported in \cite{time}.
Taking the total KARMEN1 data set of so-called 'single prong' events
(unaccompanied energy deposits without delayed coincidences) collected from 
July 1990 through December 1995, 
$N_X=89\pm 24$ events were observed in excess to the neutrino induced events 
and the cosmic background. Detailed analyses and special measurements did not 
lead to the identification of potential background sources(e.g. beam 
correlated neutrons from the spallation source, electronic effects, detector 
inefficiencies or after-pulses of the ISIS proton synchrotron) 
as origin of this excess.
  \begin{figure}[hbt]
  \centerline{\epsfig{figure=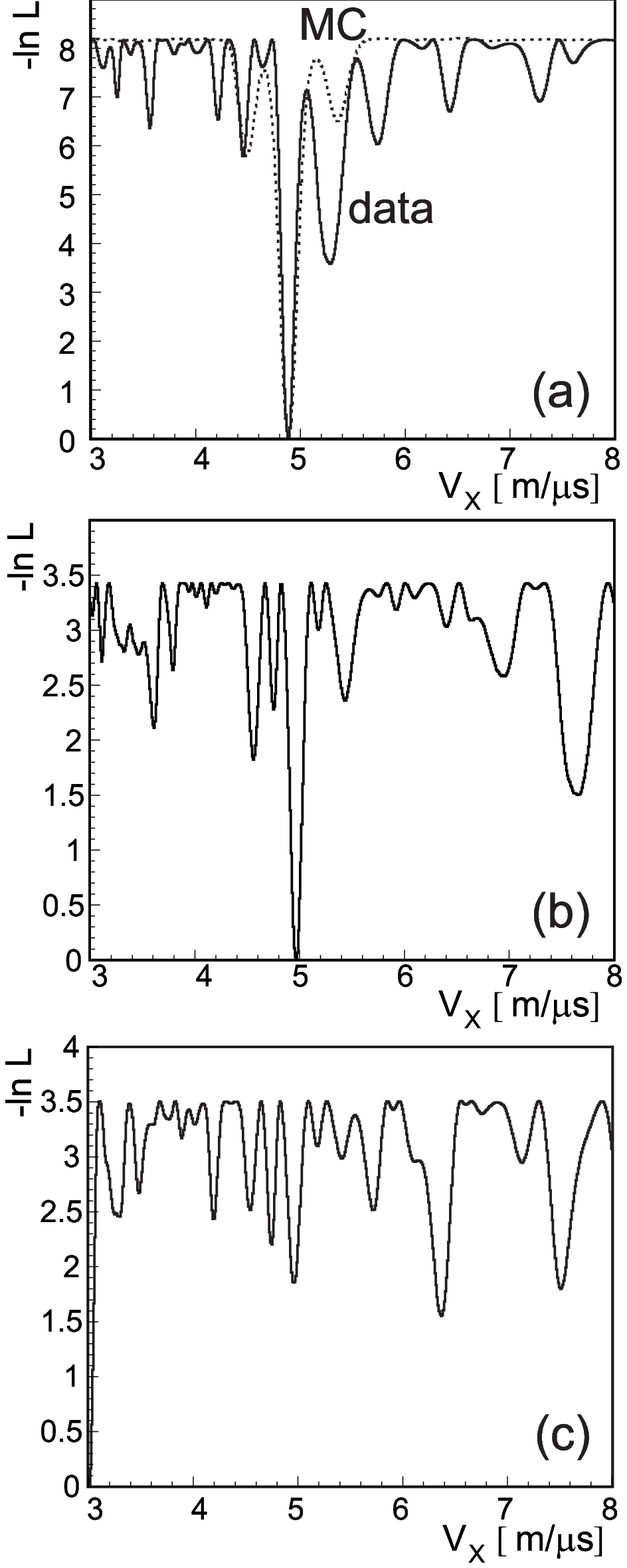,width=7cm}}
  \caption{Likelihood as a function of the velocity $v_X$ for KARMEN1(a), 
	initial KARMEN2(b) and all KARMEN2(c) data (see text for details).}
  \label{anom}
  \end{figure}
Therefore, a working hypothesis of a particle production was considered to 
get a consistent description of the origin of these events. 

The ansatz consisted of a rare
non-SM pion decay mode \pipmupx\ where $X$ denotes a massive 
($m_X\approx 33.9$\,MeV), neutral, weakly interacting, unstable particle with 
decay products detectable via electromagnetic interactions. A possible
branching ratio as small as $10^{-16}$ relative to the standard decay
\pipmup\ could have explained the observed
excess without contradiction from other experiments.
In this framework, the X particle would be produced in the ISIS target
within the pion decay double pulse structure (see Fig.~\ref{isis_nu}a), then
slowly moving ($v_X\approx 5$\,m/\us) towards the KARMEN detector at a mean
distance of $\langle d \rangle = 17.7$\,m. This would result in 
a specific double pulse structure in time and position within the 3.5\,m long
scintillation tank. The consistency of this hypothesis was tested by
a maximum likelihood analysis of individual event times and locations 
\cite{chris}. The free parameters of the likelihood function were the 
number $N_X$ of $X$ particle signatures and their velocity $v_X$.

The result of such an analysis of the KARMEN1 data (1990--95), the negative
logarithm of the likelihood $-lnL$ is shown in Figure~\ref{anom}a) as a 
function of one of the two
free parameters, $v_X$. A clear minimum at $v_X=4.89$\,m/\us\ is evident. 
The characteristic side maxima result from the double pulse structure of pion 
decays as underlined by Monte Carlo studies with high statistics 
(MC in Fig.~\ref{anom}a) of $X$ particle signals.
Hence, the working hypothesis gave a consistent description of the anomaly
in the time spectrum with a signal strength of $N_X=57\pm 25$ excess events.

One of the major objectives of KARMEN2 has been the investigation of this
hypothesis, taking full advantage of the significantly reduced level of
cosmic background.
An initial KARMEN2 data sample sufficiently large for a detailed maximum 
likelihood analysis was collected from February 1997 through February 1999.  
The corresponding maximum likelihood result $-lnL$ of these first 2 years
is given in Figure~\ref{anom}b) having its 
minimum at $v_X=4.96$\,m/\us\ with a signal strength of $N_X=17\pm 11$. Scaling
from the KARMEN1 analysis, one would have expected  $N_X^{expected}=22\pm 10$. 
Though the minimum of the function $-lnL$ was not deep enough to form a 
statistically significant signal by KARMEN2 data alone, the fitted particle 
velocity and signal strength were in good agreement with the earlier KARMEN1 
data, indicating a continuous accumulation of $X$ signatures even under 
modified experimental systematics.

However, an updated analysis of all KARMEN2 data collected so far
(February 1997 through March 2000) reveals no distortion of the time spectrum
anymore. Despite an increase of pion and 
consequently neutrino production at the ISIS target by more than 50\%,
the number of excess events in the time
spectrum between 3.1\,\us\ and 4.1\,\us\ after beam-on-target determined by
subtraction of a fitted neutrino signal (exponential decrease with 
 $\tau=2.197$\,\us) and constant background is now $N_X=11\pm 12.4$.
This allows to set an upper limit of $N_X<30$ at a 90\% 
confidence level. This has to be compared
with an expected excess of $N_X^{expected}=53\pm 14$ extrapolated from the
KARMEN1 data. In addition to the time spectrum, the twodimensional maximum
likelihood analysis shows no significant minimum of the negative logarithmic 
likelihood function at a velocity near $v_X=4.96$\,m/\us\ 
(see Fig.~\ref{anom}c). Fixing the velocity $v_X$ at the above value results
in a signal strength of $N_X=14\pm 12$ events in contradiction to the value
of $N_X^{expected}=34\pm 15$ extrapolated from the KARMEN1 
likelihood analysis.

To summarize, the time spectrum of the entire KARMEN2 data after 3 years
of data taking
is in good agreement with the expectation of neutrino induced events with an 
exponential shape superimposed on a flat cosmic induced background. There is 
no distortion in the time spectrum nor any unexplained signal in a detailed 
maximum likelihood analysis as observed for the KARMEN1 data. 
As there was no change since 1997 in data taking nor in the 
reduction and evaluation of the data, the new results of the
analysis are due to the increased data amount and statistics only. On the other
hand, the clear distortion of the KARMEN1 spectrum remains unchanged
and unexplained. While analysing the total KARMEN1+2 data set still
yields a significant minimum in $-lnL$ at $v_X=4.90$\,m/\us ,
the working hypothesis of \pipmupx\ has to be rejected on the basis of the
incompatibility of the 'signal' strength seen by KARMEN1 and KARMEN2.

KARMEN will continue to take data for another year, mainly to increase the
sensitivity of the oscillation search, but also to monitor the time spectrum
of single prong events.

\end{document}